\documentclass{article}
\usepackage{spconf,amsmath,graphicx}
\usepackage{multirow,amssymb,amsmath,amsfonts}
\usepackage{graphicx}
\usepackage{wasysym}
\usepackage{amssymb}
\usepackage{graphicx}
\usepackage{psfrag}
\usepackage{epsfig}
\usepackage{wrapfig}
\usepackage{cite}
\usepackage{blindtext}
\usepackage[table]{xcolor}
\usepackage{url}
\usepackage{float}
\usepackage{enumerate}
\usepackage{multirow}
\usepackage{slashbox}
\usepackage{caption}
\usepackage{epstopdf}
\usepackage{adjustbox}
\newcommand\tab[1][.5cm]{\hspace*{#1}}
\usepackage{array}
\usepackage[linesnumbered,ruled]{algorithm2e}
\newcolumntype{?}[1]{!{\vrule width #1}}
\usepackage{subfig}
\usepackage{booktabs}

\title{Generalization of Spoofing Countermeasures: a Case Study with ASVspoof 2015 and BTAS 2016 Corpora}

\name{Dipjyoti Paul$^{1}$, Md Sahidullah$^{2}$, Goutam Saha$^{1}$}
\address{$^{1}$Department of E \& ECE, Indian Institute of Technology Kharagpur, Kharagpur, India\\
$^{2}$School of Computing, University of Eastern Finland, Joensuu, Finland\\
e-mail: dipjyotipaul@ece.iitkgp.ernet.in, sahid@cs.uef.fi, gsaha@ece.iitkgp.ernet.in}

\begin{document}
\maketitle
\vspace{-1em}
\begin{abstract}
\vspace{-0.25em}
Voice-based biometric systems are highly prone to spoofing attacks. Recently, various countermeasures have been developed for detecting different kinds of attacks such as replay, speech synthesis (SS) and voice conversion (VC). Most of the existing studies are conducted with a specific training set defined by the evaluation protocol. However, for realistic scenarios, selecting appropriate training data is an open challenge for the system administrator. Motivated by this practical concern, this work investigates the generalization capability of spoofing countermeasures in restricted training conditions where speech from a broad attack types are left out in the training database. We demonstrate that different spoofing types have considerably different generalization capabilities. For this study, we analyze the performance using two kinds of features, mel-frequency cepstral coefficients (MFCCs) which are considered as baseline and recently proposed constant Q cepstral coefficients (CQCCs). The experiments are conducted with standard Gaussian mixture model - maximum likelihood (GMM-ML) classifier on two recently released spoofing corpora: ASVspoof 2015 and BTAS 2016 that includes cross-corpora performance analysis. Feature-level analysis suggests that static and dynamic coefficients of spectral features, both are important for detecting spoofing attacks in the real-life condition.
\end{abstract}
\vspace{-0.5em}
\begin{keywords}
Spoofing Attack, Replay Attack, ASVspoof 2015, BTAS 2016, Generalized countermeasure.
\end{keywords}
\vspace{-7pt}
\section{Introduction}
\vspace{-7pt}
\label{sec:intro}
\emph{Spoofing attacks} imitate a person's identity in order to gain illegitimate access to sensitive or protected resources. Nowadays, significant advancement in speech technology related to SS and VC techniques poses threat to speech-based biometric systems like \emph{automatic speaker verification} (ASV) systems \cite{wu2015spoofing}. \emph{Replay attacks} are another form of spoofing attack, where an adversary tries to attack a system using pre-recorded speech accumulated from target speakers \cite{lindberg1999vulnerability}. Due to the availability of high-quality, low-cost recording and playback devices, replay attacks are also a serious threat to the voice biometric systems. Several replay spoofing detection approaches such as fixed pass-phrase method, spectral ratio and modulation index were proposed in \cite{shang2010score, villalba2011detecting, villalba2011preventing}. A study on cross database evaluation was demonstrated in \cite{korshunov2016cross}.

To detect SS and VC attacks, diverse range of feature extraction methods such as \emph{mel-frequency cepstral coefficients} (MFCCs) cepstral feature \cite{wu2012detecting}, phase features \cite{de2012evaluation, sanchez2015aholab, alam2015development}, a combination of both amplitude and phase feature \cite{xiao2015spoofing}, prosodic features \cite{de2012synthetic} were reported. A concise experimental review of spoofing detection was presented in \cite{sahidullah2015comparison}. While, MFCCs are considered as the standard feature extraction techniques in speech processing, \emph{constant Q transform cepstral coefficients} (CQCCs) have shown best detection performance, especially for unknown attacks in ASVspoof 2015 corpus \cite{todisconew}. However, it was not implemented for replay attack detection.
\begin{figure*}[h!]
\vspace{-3.2em}
    \centering
    \includegraphics[height=6.5cm,width=1.23\textwidth]{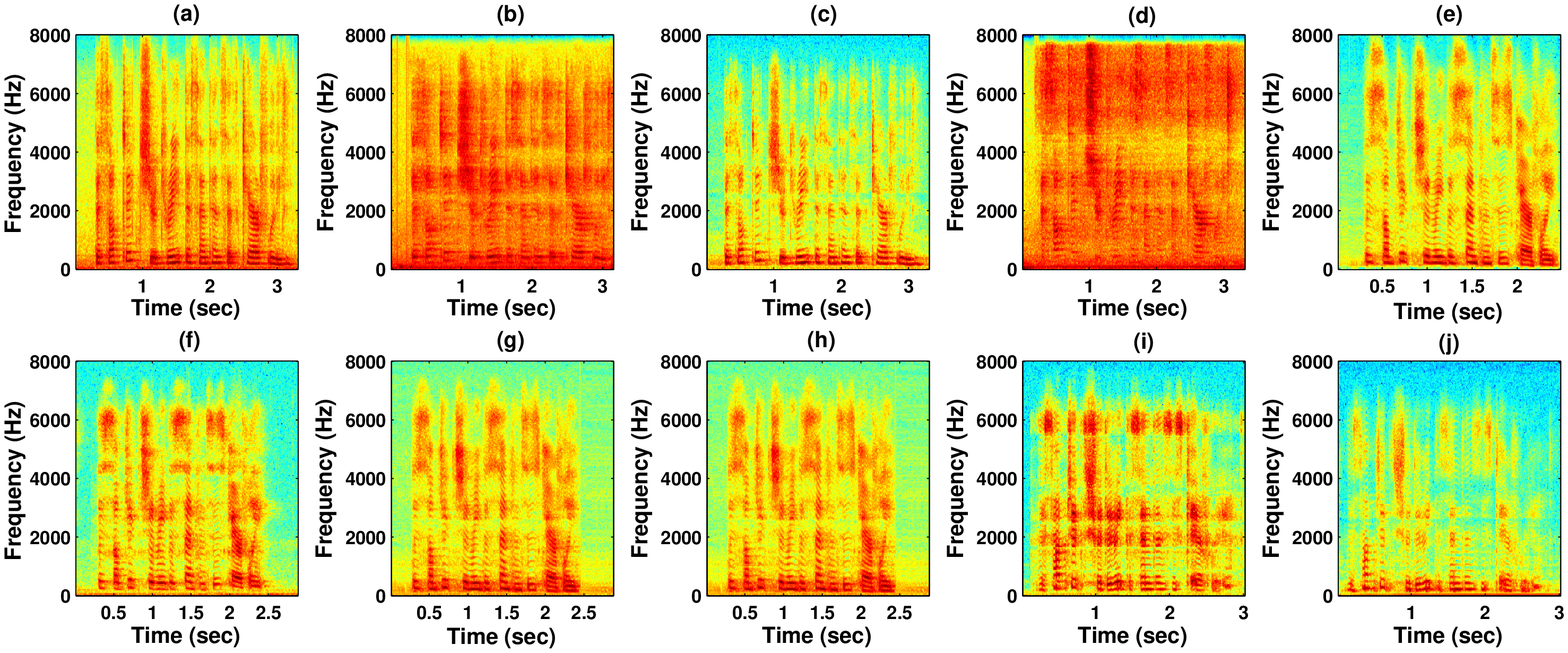}
    \vspace{-2.2em}
    \caption{\small Spectrogram of (a) genuine and replay speech signals for same sentence \textit{``The subject should read the sentences carefully"}. The replayed signals are generated by using techniques based on (b) replay laptop, (c) replay laptop high quality, (d) replay phone, (e) SS, (f) replay SS, (g) replay SS high quality, (h) VC, (i) replay VC and (j) replay VC high quality.}
  \label{figure:spectrogram plot}
  \vspace{-1.5em}
\end{figure*}

Techniques to generate voice converted speech and synthetic speech, made a rapid progress in recent times. Notable among them are \emph{joint density-Gaussian mixture model} (JD-GMM) \cite{toda2007voice}, \emph{line spectrum pairs} (LSP) \cite{saito2011one}, MARY \emph{text-to-speech synthesis} (MARY-TTS) \cite{schroder2003german}, \emph{hidden Markov model} (HMM) based TTS \cite{yamagishi2009analysis} etc. It is not practically possible to anticipate the kind of SS and VC attack all the time to include those types of speeches in the training database. At the same time, it is expected that detection performance will degrade if similar kinds of data are unavailable in the training corpus. The previous studies on spoofing detection do not focus on \emph{attack dependency} which is the central theme of this work. There are some results generated in recent spoofing challenges with \emph{unknown attack} types but no exhaustive study is done that can lead to \emph{generalization} ability of certain training schemes over other for a range of unknown attacks.

In this work, we did a systematic study of attack dependency to discover corresponding generalization ability. We demonstrate the result using conventional MFCCs and newly proposed CQCCs features on GMM-maximum likelihood (GMM-ML) framework. It is found that GMM-ML as a classifier is better suited for spoofing detection task \cite{hanilcci2015classifiers}. We have experimented on two recent databases: ASVspoof 2015, developed as a part of \emph{Automatic Speaker Verification Spoofing and Countermeasure Challenge} \cite{wu2015asvspoof} and BTAS 2016 corpus in \emph{Speaker Anti-spoofing Competition} \cite{korshunovoverview}. BTAS 2016 introduces more realistic replay attacks compared to ASVspoof database. Our study shows the generalization ability of one countermeasure over the other.
\vspace{-10pt}
\section{Generalization Framework}
\vspace{-8pt}

\begin{figure}[h!]
  \centering
  \includegraphics[height=4.5cm,width=0.50\textwidth]{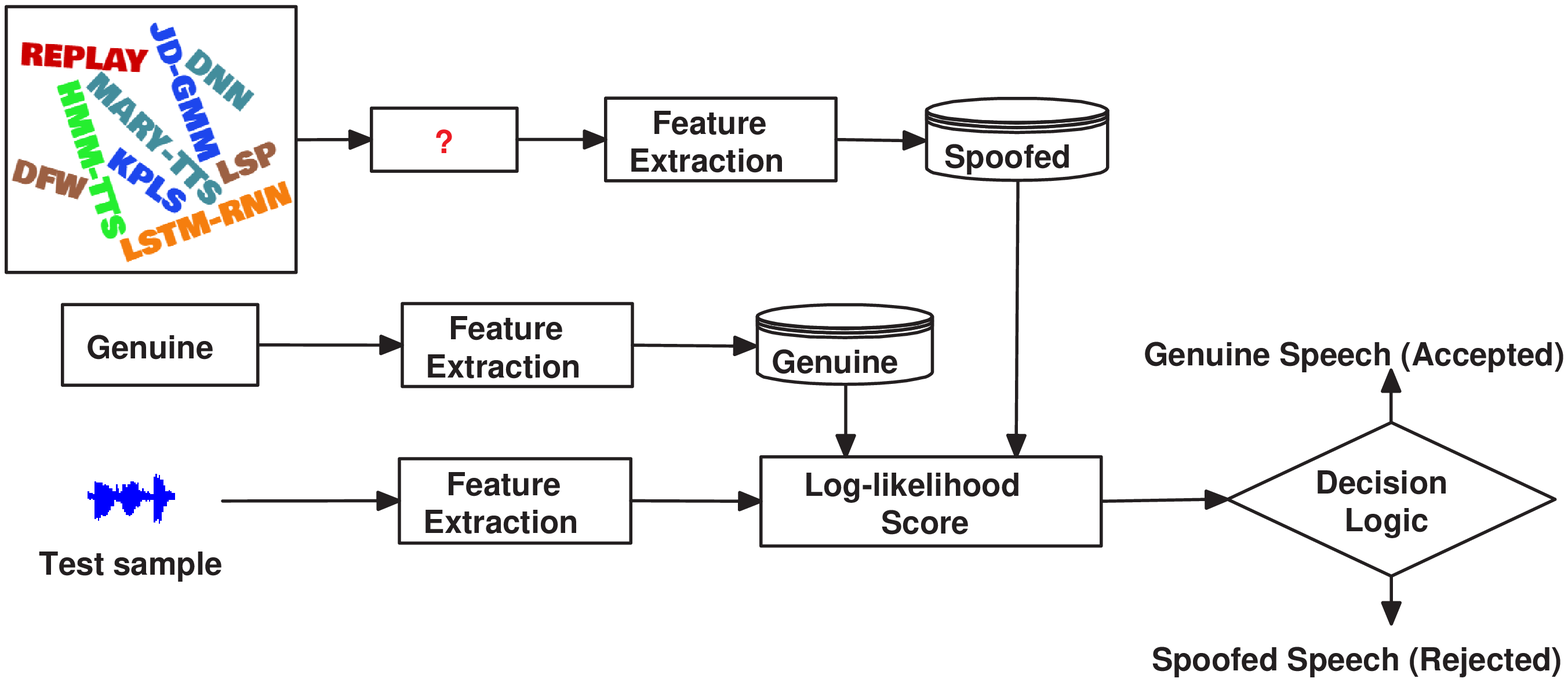}
  \caption{A speech-based countermeasure system.}
\label{spectrogram}
\vspace{-1em}
\end{figure}

Figure \ref{figure:spectrogram plot} illustrates the spectral characteristics of spoofed signals for diverse attacks. \emph{Generalized countermeasure} refers to the ability to overcome the attack dependency in the detection process. This dependency signifies the types of attacks that are best represented by a similar pattern in the attack space. It involves a prior knowledge of attack type, which is not a realistic assumption in all the cases. Therefore, the countermeasure system needs to be robust enough to detect an attack even though that type of attack data is not used for training the model. Fig. \ref{spectrogram} describes the functional block diagram of a generalized countermeasure framework where we find which kind of training has greater generalization ability.
\vspace{-5pt}

Initially, we train the models using all types of replay (i.e., genuine, SS and VC samples) and synthetic (SS and VC samples) attacks. Then, we study the impact when one type of spoofing data is not used for modeling the spoofed data.
\vspace{-13pt}
\section{Experimental Setup}
\vspace{-8pt}
\subsection{Database Description}
\vspace{-5pt}
\tab \textbf{ASVspoof 2015:} ASVspoof database is created to assess ten different types of SS and VC synthetic speech samples namely S1 to S10 \cite{wu2015asvspoof}. It includes both the known attacks (S1-S5) and the unknown attacks (S6-S10).

\textbf{BTAS 2016:} BTAS database contains genuine and different kinds of replay attacks where genuine, SS and VC speech samples were played back using high-quality devices. Two new replay unknown attacks (R9 and R10) are introduced in the evaluation data to make it more challenging. The statistics regarding types of attacks and the number of utterances for each dataset are presented in Table \ref{database}.
\vspace{-0.8em}
\begin{table}[h!]
\centering
\scriptsize
\setlength{\tabcolsep}{3.2pt}
\renewcommand{\arraystretch}{1.2}
\caption{\small Number of utterances in BTAS 2016 database. LP: laptop, HQ: high quality speaker, PH1: Samsung Galaxy S4 phone, PH2: iPhone 3GS and ‘PH3’ is iPhone 6S.}
\vspace{-0.3cm}
\label{database}
\begin{tabular}{|c|c|c|ccc|}
\hline
\multicolumn{3}{|c|}{Types}                                                                               & Training & Development & Evaluation \\ \hline
\multicolumn{3}{|c|}{Genuine}                                                                             & 4973     & 4995        & 5576       \\ \hline
\multirow{6}{*}{Replay} & Replay LP LP            & R1  & 700      & 700         & 800        \\
                                                                          & Replay LP HQ LP         & R2  & 700      & 700         & 800        \\
                                                                          & Replay PH1 LP           & R3  & 700      & 700         & 800        \\
                                                                          & Replay PH2 LP           & R4  & 700      & 700         & 800        \\
                                                                          & Replay PH2 PH3     & R9  & -        & -           & 800        \\
                                                                          & Replay LP PH2 PH3  & R10 & -        & -           & 800        \\ \hline
\multirow{2}{*}{SS}      & SS LP LP           & R5  & 490      & 490         & 560        \\
                                                                          & SS LP HQ LP        & R6  & 490      & 490         & 560        \\ \hline
\multirow{2}{*}{VC}      & VC LP LP          & R7  & 17400    & 17400       & 19500      \\
                                                                          & VC LP HQ LP        & R8  & 17400    & 17400       & 19500      \\ \hline
\end{tabular}
\vspace{-0.3cm}
\end{table}
\vspace{-7pt}
\begin{table*}[h!]
\centering
\scriptsize
\setlength{\tabcolsep}{5pt}
\renewcommand{\arraystretch}{0.98}
\caption{\small Performance (in \% of EER) for MFCC and CQCC features on BTAS 2016 development data. The corresponding class of attacks that are not considered in the training are highlighted.}
\vspace{-0.3cm}
\label{dev}
\begin{tabular}{|c|ccc|c|cccccccc|cccc|}
\hline
                           & \multicolumn{3}{c|}{Train}                                                                 &                            &                               &                               &                               &                               &                              &                              &                                                     &                                                   &\multicolumn{4}{c|}{Average}                      \\ \cline{2-4}\cline{14-17}
\multirow{-2}{*}{Features} & Replay               & SS                   & VC                    & \multirow{-2}{*}{Type}     & \multirow{-2}{*}{R1}          & \multirow{-2}{*}{R2}          & \multirow{-2}{*}{R3}          & \multirow{-2}{*}{R4}          & \multirow{-2}{*}{R5}         & \multirow{-2}{*}{R6}         & \multirow{-2}{*}{R7}                                & \multirow{-2}{*}{R8}              &Replay&SS&VC                  & All \\ \hline
                           &                              &                              &                              & Static                     & 0.14                          & 0.61                          & 0.09                          & 0.00                          & 0.00                         & 0.84                         & 0.00                                                & 0.02                                               &0.21&0.42&0.01 & 0.21                  \\
                           &                              &                              &                              & Static+$\Delta \Delta^{2}$ & 0.34                          & 2.74                          & 0.00                          & 0.00                          & 0.00                         & 0.57                         & 0.00                                                & 0.01                                               &0.77&0.29&0.01 & 0.46                  \\
                           & \multirow{-3}{*}{\checkmark} & \multirow{-3}{*}{\checkmark} & \multirow{-3}{*}{\checkmark} & $\Delta \Delta^{2}$        & 19.50                         & 41.35                         & 28.62                         & 28.67                         & 1.33                         & 0.86                         & 0.00                                                & 0.01                                               &29.54&1.10&0.01 & 15.04                 \\ \cline{2-17}
                           &                              &                              &                              & Static                     & \cellcolor[HTML]{C0C0C0}4.91  & \cellcolor[HTML]{C0C0C0}5.92  & \cellcolor[HTML]{C0C0C0}30.33 & \cellcolor[HTML]{C0C0C0}24.56 & 0.00                         & 1.07                         & 0.01                                                & 0.01                                            &16.43&0.54&0.01    & \textbf{8.35}         \\
                           &                              &                              &                              & Static+$\Delta \Delta^{2}$ & \cellcolor[HTML]{C0C0C0}4.73  & \cellcolor[HTML]{C0C0C0}6.98  & \cellcolor[HTML]{C0C0C0}30.51 & \cellcolor[HTML]{C0C0C0}26.29 & 0.00                         & 0.65                         & 0.01                                                & 0.00                                             &17.13&0.33&0.01   & 8.65                  \\
                           & \multirow{-3}{*}{$\times$}   & \multirow{-3}{*}{\checkmark} & \multirow{-3}{*}{\checkmark} & $\Delta \Delta^{2}$        & \cellcolor[HTML]{C0C0C0}25.44 & \cellcolor[HTML]{C0C0C0}44.39 & \cellcolor[HTML]{C0C0C0}34.34 & \cellcolor[HTML]{C0C0C0}35.52 & 1.33                         & 1.10                         & 0.00                                                & 0.01                                               &34.92&1.23&0.01 & 17.77                 \\ \cline{2-17}
                           &                              &                              &                              & Static                     & 0.27                          & 0.63                          & 0.00                          & 0.00                          & \cellcolor[HTML]{C0C0C0}0.00 & \cellcolor[HTML]{C0C0C0}1.93 & 0.00                                                & 0.02                                               &0.23&0.97&0.01 & 0.36                  \\
                           &                              &                              &                              & Static+$\Delta \Delta^{2}$ & 0.31                          & 3.50                          & 0.00                          & 0.00                          & \cellcolor[HTML]{C0C0C0}0.00 & \cellcolor[HTML]{C0C0C0}2.07 & 0.00                                                & 0.01                                               &0.95&1.04&0.01 & 0.74                  \\
                           & \multirow{-3}{*}{\checkmark} & \multirow{-3}{*}{$\times$}   & \multirow{-3}{*}{\checkmark} & $\Delta \Delta^{2}$        & 19.63                         & 40.88                         & 27.35                         & 27.99                         & \cellcolor[HTML]{C0C0C0}2.39 & \cellcolor[HTML]{C0C0C0}1.62 & 0.00                                                & 0.01                                               &28.96&2.01&0.01 & 14.98                 \\ \cline{2-17}
                           &                              &                              &                              & Static                     & 0.00                          & 0.30                          & 0.00                          & 0.00                          & 0.00                         & 0.04                         & \cellcolor[HTML]{C0C0C0}0.32                        & \cellcolor[HTML]{C0C0C0}2.13                       &0.15&0.02&1.23 & 0.35                  \\
                           &                              &                              &                              & Static+$\Delta \Delta^{2}$ & 0.04                          & 0.08                          & 0.00                          & 0.00                          & 0.00                         & 0.02                         & \cellcolor[HTML]{C0C0C0}0.55                        & \cellcolor[HTML]{C0C0C0}0.85                      &0.03&0.01&0.70  & 0.19                  \\
\multirow{-12}{*}{MFCC}    & \multirow{-3}{*}{\checkmark} & \multirow{-3}{*}{\checkmark} & \multirow{-3}{*}{$\times$}   & $\Delta \Delta^{2}$        & 2.44                          & 24.05                         & 5.80                          & 2.69                          & 0.14                         & 0.85                         & \cellcolor[HTML]{C0C0C0}3.60                        & \cellcolor[HTML]{C0C0C0}6.30                       &8.75&0.50&4.95 & 5.73                  \\ \hline
                           &                              &                              &                              & Static                     & 0.00                          & 0.00                          & 0.00                          & 0.00                          & 0.00                         & 0.00                         & 0.00                                                & 0.00                                              &0.00&0.00&0.00  & \textbf{0.00}         \\
                           &                              &                              &                              & Static+$\Delta \Delta^{2}$ & 0.00                          & 0.00                          & 0.00                          & 0.00                          & 0.00                         & 0.00                         & 0.00                                                & 0.00                                               &0.00&0.00&0.00 & 0.00                  \\
                           & \multirow{-3}{*}{\checkmark} & \multirow{-3}{*}{\checkmark} & \multirow{-3}{*}{\checkmark} & $\Delta \Delta^{2}$        & 20.04                         & 6.43                          & 25.22                         & 41.05                         & 1.10                         & 0.11                         & 0.09                                                & 0.18                                               &23.19&0.61&0.14 & 11.78                 \\ \cline{2-17}
                           &                              &                              &                              & Static                     & \cellcolor[HTML]{C0C0C0}1.27  & \cellcolor[HTML]{C0C0C0}0.00  & \cellcolor[HTML]{C0C0C0}47.71 & \cellcolor[HTML]{C0C0C0}40.84 & 0.00                         & 0.00                         & 0.00                                                & 0.00                                              &22.46&0.00&0.00  & 11.23                 \\
                           &                              &                              &                              & Static+$\Delta \Delta^{2}$ & \cellcolor[HTML]{C0C0C0}8.25  & \cellcolor[HTML]{C0C0C0}0.00  & \cellcolor[HTML]{C0C0C0}49.18 & \cellcolor[HTML]{C0C0C0}44.75 & 0.00                         & 0.00                         & 0.00                                                & 0.00                                              &25.55&0.00&0.00  & 12.77                 \\
                           & \multirow{-3}{*}{$\times$}   & \multirow{-3}{*}{\checkmark} & \multirow{-3}{*}{\checkmark} & $\Delta \Delta^{2}$        & \cellcolor[HTML]{C0C0C0}25.07 & \cellcolor[HTML]{C0C0C0}9.36  & \cellcolor[HTML]{C0C0C0}33.72 & \cellcolor[HTML]{C0C0C0}43.77 & 1.03                         & 0.23                         & 0.09                                                & 0.19                                                &27.98&0.63&0.14 & 14.18                 \\ \cline{2-17}
                           &                              &                              &                              & Static                     & 0.00                          & 0.00                          & 0.00                          & 0.00                          & \cellcolor[HTML]{C0C0C0}0.00 & \cellcolor[HTML]{C0C0C0}0.00 & 0.00                                                & 0.00                                              &0.00&0.00&0.00  & \textbf{0.00}         \\
                           &                              &                              &                              & Static+$\Delta \Delta^{2}$ & 0.00                          & 0.00                          & 0.00                          & 0.00                          & \cellcolor[HTML]{C0C0C0}0.00 & \cellcolor[HTML]{C0C0C0}0.00 & 0.00                                                & 0.00                                               &0.00&0.00&0.00 & 0.00                  \\
                           & \multirow{-3}{*}{\checkmark} & \multirow{-3}{*}{$\times$}   & \multirow{-3}{*}{\checkmark} & $\Delta \Delta^{2}$        & 20.31                         & 6.48                          & 25.54                         & 40.93                         & \cellcolor[HTML]{C0C0C0}2.26 & \cellcolor[HTML]{C0C0C0}0.20 & 0.08                                                & 0.18                                              &23.32&1.23&0.13  & 12.00                 \\ \cline{2-17}
                           &                              &                              &                              & Static                     & 0.00                          & 0.00                          & 0.00                          & 0.00                          & 0.00                         & 0.00                         & \cellcolor[HTML]{C0C0C0}{\color[HTML]{000000} 0.00} & \cellcolor[HTML]{C0C0C0}{\color[HTML]{000000} 0.00} &0.00&0.00&0.00& \textbf{0.00}         \\
                           &                              &                              &                              & Static+$\Delta \Delta^{2}$ & 0.00                          & 0.00                          & 0.00                          & 0.00                          & 0.00                         & 0.00                         & \cellcolor[HTML]{C0C0C0}{\color[HTML]{000000} 0.00} & \cellcolor[HTML]{C0C0C0}{\color[HTML]{000000} 0.00} &0.00&0.00&0.00& 0.00                  \\
\multirow{-12}{*}{CQCC}    & \multirow{-3}{*}{\checkmark} & \multirow{-3}{*}{\checkmark} & \multirow{-3}{*}{$\times$}   & $\Delta \Delta^{2}$        & 10.52                         & 2.71                          & 18.08                         & 29.93                         & 0.41                         & 0.30                         & \cellcolor[HTML]{C0C0C0}{\color[HTML]{000000} 2.46} & \cellcolor[HTML]{C0C0C0}{\color[HTML]{000000} 2.89} &15.31&0.36&2.68& 8.37                  \\ \hline
\end{tabular}
\vspace{-0.3cm}
\end{table*}
\begin{table*}[h!]
\centering
\scriptsize
\setlength{\tabcolsep}{5pt}
\renewcommand{\arraystretch}{0.98}
\caption{\small Performance (in \% of EER) for MFCC and CQCC (static) features on BTAS 2016 evaluation data. The corresponding class of attacks that are not considered in the training systems are highlighted.}
\vspace{-0.3cm}
\label{eval}
\begin{tabular}{|c|ccc|cccccccccc|cccc|}
\hline
                           & \multicolumn{3}{c|}{Train}                                                                          &                               &                               &                               &                               &                              &                              &                              &                              & \multicolumn{1}{c}{}                             & \multicolumn{1}{c|}{}           &\multicolumn{4}{c|}{Average}               \\ \cline{2-4}\cline{15-18}
\multirow{-2}{*}{Features} & \multicolumn{1}{c}{Replay} & \multicolumn{1}{c}{SS} & \multicolumn{1}{c|}{VC} & \multirow{-2}{*}{R1}          & \multirow{-2}{*}{R2}          & \multirow{-2}{*}{R3}          & \multirow{-2}{*}{R4}          & \multirow{-2}{*}{R5}         & \multirow{-2}{*}{R6}         & \multirow{-2}{*}{R7}         & \multirow{-2}{*}{R8}         & \multicolumn{1}{c}{\multirow{-2}{*}{R9}}         & \multicolumn{1}{c|}{\multirow{-2}{*}{R10}}     &Replay   &SS &VC      & All \\ \hline
                           & \checkmark                         & \checkmark                    & \checkmark                     & 1.11                          & 8.26                          & 0.01                          & 0.14                          & 0.06                         & 3.67                         & 0.03                         & 1.68                         & \multicolumn{1}{c}{17.87}                         & \multicolumn{1}{c|}{10.85}                              &6.37&1.87&0.86&4.37  \\
                           & $\times$                         &  \checkmark                     & \checkmark                       & \cellcolor[HTML]{C0C0C0}6.61  & \cellcolor[HTML]{C0C0C0}16.12 & \cellcolor[HTML]{C0C0C0}32.89 & \cellcolor[HTML]{C0C0C0}24.38 & 0.14                         & 5.34                         & 0.06                         & 2.88                         & \multicolumn{1}{l}{\cellcolor[HTML]{C0C0C0}30.50} & \multicolumn{1}{c|}{\cellcolor[HTML]{C0C0C0}40.89} &25.23 &2.74&1.47&15.98                  \\
                           & \checkmark                           & $\times$                    & \checkmark                       & 1.42                          & 8.41                          & 0.02                          & 0.11                          & \cellcolor[HTML]{C0C0C0}0.73 & \cellcolor[HTML]{C0C0C0}5.80 & 0.04                         & 1.84                         & \multicolumn{1}{c}{18.21}                         & \multicolumn{1}{c|}{10.23}                         &6.40&3.27&0.94& 4.68                  \\
\multirow{-4}{*}{MFCC}     & \checkmark                           & \checkmark                      & $\times$                     & 0.22                          & 2.31                          & 0.00                             & 0.00                             & 0.00                            & 0.16                         & \cellcolor[HTML]{C0C0C0}3.06 & \cellcolor[HTML]{C0C0C0}4.90 & \multicolumn{1}{l}{19.47}                         & \multicolumn{1}{c|}{6.59}                        &4.77&0.08&3.98 & 3.67                 \\ \hline

                           & \checkmark                         & \checkmark                    & \checkmark                     & 0.00                             & 0.04                          & 0.00                             & 0.00                             & 0.00                            & 0.00                            & 0.00                            & 0.00                            & 7.56                                                & 0.00                                               &1.27&0.00&0.00 & \textbf{0.76}         \\
                           & $\times$                         & \checkmark                      & \checkmark                       & \cellcolor[HTML]{C0C0C0}7.20  & \cellcolor[HTML]{C0C0C0}0.43  & \cellcolor[HTML]{C0C0C0}48.79 & \cellcolor[HTML]{C0C0C0}43.96 & 0.00                            & 0.00                            & 0.00                            & 0.01                         & \cellcolor[HTML]{C0C0C0}10.02                        & \cellcolor[HTML]{C0C0C0}32.76                     &23.86&0.00&0.10& \textbf{14.32}                 \\
                           & \checkmark                           & $\times$                    & \checkmark                       & 0.00                             & 0.00                             & 0.00                             & 0.00                             & \cellcolor[HTML]{C0C0C0}0.19 & \cellcolor[HTML]{C0C0C0}0.00    & 0.00                            & 0.00                            & 13.08                                                & 0.03                                               &2.19&0.10&0.00 & \textbf{1.33}         \\
\multirow{-4}{*}{CQCC}     & \checkmark                           & \checkmark                      & $\times$                     & 0.00                             & 0.00                             & 0.00                             & 0.00                             & 0.00                            & 0.00                            & \cellcolor[HTML]{C0C0C0}0.25 & \cellcolor[HTML]{C0C0C0}0.01 & 12.19                                             & 0.03                                              &2.04&0.00&0.13 & \textbf{1.25}
\\ \hline
\end{tabular}
\vspace{-0.35cm}
\end{table*}

\vspace{-10pt}
\subsection{Feature Extraction Techniques}
\vspace{-8pt}
\tab \textbf{Mel-frequency cepstral coefficients (MFCCs):} MFCC \cite{davis1980comparison} feature utilizes mel-scale based triangular filter bank. The power spectrum is integrated using overlapping band-pass filters in the triangular filterbank. We use the configuration reported in \cite{sahidullah2015comparison}.

\textbf{Constant Q cepstral coefficients (CQCCs):} The constant Q transform (CQT) gives a higher frequency resolution in lower frequencies and a greater temporal resolution in the higher frequency region. A spline interpolation method is applied to resample the geometric frequency scale into a uniform linear scale in order to apply linearly spaced DCT coefficients for CQCC cepstral feature computation \cite{todisconew}.

CQCC feature is implemented with maximum frequency $(f_{max}=4 Khz)$ and minimum frequency of $(f_{min}=15 Hz)$. The number of bins per octave is assigned to 96. Speech activity detector (SAD) is not employed as non-speech frames could be helpful for spoofing detection.
\vspace{-10pt}
\subsection{Classifier and Performance Evaluation}
\vspace{-5pt}
We employ GMM-ML classifier for spoofing detection. Two target models $\boldsymbol{\lambda}_{n}$ and $\boldsymbol{\lambda}_{s}$  are created from natural and spoofed speech data respectively \cite{dpaul15}. The log-likelihood score is calculated as, $ \Lambda(\mathbf{X})=\mathcal{L}(\textbf{X}|\boldsymbol{\lambda}_{n})-\mathcal{L}(\mathbf{X}|\boldsymbol{\lambda}_{s}),$ where $\mathbf{X}=\{\mathbf{x}_{1},\ldots, \mathbf{x}_{T}\}$ is the feature matrix of the test utterance, $T$ is the number of frames and $\mathcal{L} (\textbf{X}|\boldsymbol{\lambda})$ is the average log-likelihood of \textbf{X} given GMM model $\boldsymbol{\lambda}$. We train GMMs with 10 iterations of expectation-maximization (EM) algorithm and 512 mixture components.

Equal error rate (EER) is used as the performance metric to evaluate spoofing attack detection. We use BOSARIS toolkit \cite{brummer2013bosaris} to calculate the EER using receiver operating characteristics convex hull (ROCCH) method. 
\begin{table*}[h!]
\renewcommand{\arraystretch}{0.98}
\centering
\scriptsize
\caption{ \small Same as Table \ref{dev} but for ASVspoof 2015 evaluation.}
\vspace{-0.3cm}
\label{asvspoof_eval}
\begin{tabular}{|c|cc|c|cccccccccc|ccc|}
\hline
                           & \multicolumn{2}{c|}{Train}                                  &                        &                               &                               &                              &                              &                               &                               &                               &                              &                               &                             &\multicolumn{3}{c|}{Average}                      \\ \cline{2-3}\cline{15-17}
\multirow{-2}{*}{Features} & SS                           & VC                           & \multirow{-2}{*}{Type} & \multirow{-2}{*}{S1}          & \multirow{-2}{*}{S2}          & \multirow{-2}{*}{S3}         & \multirow{-2}{*}{S4}         & \multirow{-2}{*}{S5}          & \multirow{-2}{*}{S6}          & \multirow{-2}{*}{S7}          & \multirow{-2}{*}{S8}         & \multirow{-2}{*}{S9}          & \multirow{-2}{*}{S10}         &SS & VC & All\\ \hline
                           &                              &                              & Static                 & 1.54                          &  7.54                             & 0.00                         & 0.00                         & 6.80                          & 5.33                          & 2.25                          & 0.04                         & 2.12                          & 26.66                         &8.89&3.66&5.23                     \\
                           & \multirow{-2}{*}{\checkmark} & \multirow{-2}{*}{\checkmark} & $\Delta \Delta^{2}$    & 0.09                          &  1.46                             & 0.00                         & 0.00                         & 0.36                          & 0.30                          & 0.02                          & 0.03                         & 0.02                          & 19.45                        &6.48&0.33& 2.17             \\ \cline{2-17}
                           &                              &                              & Static                 & 1.64                          & 7.35                          & \cellcolor[HTML]{C0C0C0}0.32 & \cellcolor[HTML]{C0C0C0}0.35 & 5.93                          & 5.07                          & 2.47                          & 0.07                         & 2.67                          & \cellcolor[HTML]{C0C0C0}30.68 &10.45&3.60&5.66              \\
                           & \multirow{-2}{*}{$\times$}   & \multirow{-2}{*}{\checkmark} & $\Delta \Delta^{2}$    & 0.00                          & 1.07                          & \cellcolor[HTML]{C0C0C0}0.30 & \cellcolor[HTML]{C0C0C0}0.28 & 0.25                          & 0.25                          & 0.01                          & 0.17                         & 0.01                          & \cellcolor[HTML]{C0C0C0}22.08 &7.55&0.25&2.44            \\ \cline{2-17}
                           &                              &                              & Static                 & \cellcolor[HTML]{C0C0C0}19.46 & \cellcolor[HTML]{C0C0C0}35.82 & 0.00                         & 0.00                         & \cellcolor[HTML]{C0C0C0}35.16 & \cellcolor[HTML]{C0C0C0}34.41 & \cellcolor[HTML]{C0C0C0}26.98 & \cellcolor[HTML]{C0C0C0}0.26 & \cellcolor[HTML]{C0C0C0}24.54 & 8.97                          &2.99 &25.23&18.56                      \\
\multirow{-6}{*}{MFCC}     & \multirow{-2}{*}{\checkmark} & \multirow{-2}{*}{$\times$}   & $\Delta \Delta^{2}$    & \cellcolor[HTML]{C0C0C0}40.96 & \cellcolor[HTML]{C0C0C0}29.02 & 0.00                         & 0.00                         & \cellcolor[HTML]{C0C0C0}11.21 & \cellcolor[HTML]{C0C0C0}11.39 & \cellcolor[HTML]{C0C0C0}2.18  & \cellcolor[HTML]{C0C0C0}0.19 & \cellcolor[HTML]{C0C0C0}2.57  & 26.23                         &8.74&13.93& 12.38             \\ \hline
                           &                              &                              & Static                 & 0.02                          & 0.44                          & 0.00                         & 0.00                         & 1.54                          & 1.04                          & 0.09                          & 0.15                         & 0.13                          & 19.11                         &6.37&0.49& 2.25             \\
                           & \multirow{-2}{*}{\checkmark} & \multirow{-2}{*}{\checkmark} & $\Delta \Delta^{2}$    & 0.02                          & 0.31                          & 0.01                         & 0.03                         & 0.27                          & 0.25                          & 0.12                          & 2.29                         & 0.15                          & 0.94                          &0.33 &0.49&  \textbf{0.44}                   \\ \cline{2-17}
                           &                              &                              & Static                 & 0.03                          & 1.43                          & \cellcolor[HTML]{C0C0C0}0.10 & \cellcolor[HTML]{C0C0C0}0.07 & 1.43                          & 1.19                          & 0.08                          & 0.59                         & 0.14                          & \cellcolor[HTML]{C0C0C0}21.77 &7.31 &0.70& 2.68                     \\
                           & \multirow{-2}{*}{$\times$}   & \multirow{-2}{*}{\checkmark} & $\Delta \Delta^{2}$    & 0.01                          & 0.08                          & \cellcolor[HTML]{C0C0C0}4.17 & \cellcolor[HTML]{C0C0C0}3.86 & 0.08                          & 0.12                          & 0.07                          & 5.93                         & 0.12                          & \cellcolor[HTML]{C0C0C0}0.72  &2.92&0.92& \textbf{1.52}                    \\ \cline{2-17}
                           &                              &                              & Static                 & \cellcolor[HTML]{C0C0C0}2.68  & \cellcolor[HTML]{C0C0C0}16.24 & 0.00                         & 0.00                         & \cellcolor[HTML]{C0C0C0}25.74 & \cellcolor[HTML]{C0C0C0}22.09 & \cellcolor[HTML]{C0C0C0}10.56 & \cellcolor[HTML]{C0C0C0}0.05 & \cellcolor[HTML]{C0C0C0}12.22 & 6.05                          &2.02&12.80& 9.56             \\
\multirow{-6}{*}{CQCC}     & \multirow{-2}{*}{\checkmark} & \multirow{-2}{*}{$\times$}   & $\Delta \Delta^{2}$    & \cellcolor[HTML]{C0C0C0}26.59 & \cellcolor[HTML]{C0C0C0}7.86  & 0.01                         & 0.03                         & \cellcolor[HTML]{C0C0C0}7.86  & \cellcolor[HTML]{C0C0C0}7.85  & \cellcolor[HTML]{C0C0C0}1.90  & \cellcolor[HTML]{C0C0C0}2.09 & \cellcolor[HTML]{C0C0C0}2.96  & 35.20                         &11.75   &8.16& \textbf{9.24}                  \\ \hline
\end{tabular}
\vspace{-0.3cm}
\end{table*}

\begin{table*}[h!]
\renewcommand{\arraystretch}{0.98}
\centering
\scriptsize
\caption{ \small Cross corpora evaluation performance (in \% of EER) on BTAS 2016 evaluation data trained using ASVspoof 2015 training dataset.}
\vspace{-0.25cm}
\label{crosscorpora_eval}
\begin{tabular}{|c|cc|c|cccccccccc|cccc|}
\hline
                           & \multicolumn{2}{c|}{Train}           &                        &                      &                      &                      &                      &                               &                               &                               &                               &                      &                 &\multicolumn{4}{c|}{Average}                       \\ \cline{2-3}\cline{15-18}
\multirow{-2}{*}{Features} & SS                           & VC                           & \multirow{-2}{*}{Type} & \multirow{-2}{*}{R1} & \multirow{-2}{*}{R2} & \multirow{-2}{*}{R3} & \multirow{-2}{*}{R4} & \multirow{-2}{*}{R5}          & \multirow{-2}{*}{R6}          & \multirow{-2}{*}{R7}          & \multirow{-2}{*}{R8}          & \multirow{-2}{*}{R9} & \multirow{-2}{*}{R10}& Replay & SS & VC & All \\ \hline
                           &                              &                              & Static                 & 46.91                & 50.00                   & 50.00                   & 50.00                   & 46.68 & 49.94 & 42.47 & 49.28 & 46.05                & 50.00                &48.83&48.31&45.88 & 48.13                 \\
                           & \multirow{-2}{*}{\checkmark} & \multirow{-2}{*}{\checkmark} & $\Delta \Delta^{2}$    & 49.93                & 34.39                & 49.88                & 49.82                & 34.31 & 4.92  & 3.97  & 4.80  & 44.62                 & 49.87                 &46.42&19.62&4.39 & \textbf{32.65}        \\ \cline{2-18}
                           &                              &                              & Static                 & 46.54                & 50.00                   & 50.00                   & 50.00                   & \cellcolor[HTML]{C0C0C0}46.53 & \cellcolor[HTML]{C0C0C0}49.97 & 42.77                         & 49.44                         & 45.90                & 49.16                &48.60&48.25&46.11 & 48.03        \\
                           & \multirow{-2}{*}{$\times$} & \multirow{-2}{*}{\checkmark}   & $\Delta \Delta^{2}$    & 49.94                   & 35.39                & 49.61                & 49.19                   & \cellcolor[HTML]{C0C0C0}39.74 & \cellcolor[HTML]{C0C0C0}7.42  & 3.20                         & 3.29                         & 47.89                & 49.85              &46.98&23.56&3.25   & \textbf{33.55}        \\ \cline{2-18}
                           &                              &                              & Static                 & 50.00                   & 49.98                & 43.23                & 47.96                & 50.00                            & 49.99                         & \cellcolor[HTML]{C0C0C0}50    & \cellcolor[HTML]{C0C0C0}46.14 & 50                   & 49.97               &48.52&50&48.07  & 48.73                 \\
\multirow{-6}{*}{MFCC}     & \multirow{-2}{*}{\checkmark}   & \multirow{-2}{*}{$\times$}  & $\Delta \Delta^{2}$    & 50.00                & 47.42                & 49.97                & 50.00                & 20.03                         & 3.26                          & \cellcolor[HTML]{C0C0C0}14.32  & \cellcolor[HTML]{C0C0C0}10.67  & 50.00                 & 49.98                 &49.56&11.65&12.50 & \textbf{34.57}        \\ \hline
                           &                              &                              & Static                 & 48.50                & 49.96                & 38.68                & 44.14                & 41.39 & 49.88 & 49.94 & 49.16 & 49.96                & 45.21               &46.08&45.64&49.55  & 46.68        \\
                           & \multirow{-2}{*}{\checkmark} & \multirow{-2}{*}{\checkmark} & $\Delta \Delta^{2}$    & 47.67                & 36.43                & 49.68                & 50                   & 19.42 & 17.29 & 18.11 & 19.97 & 46.64                & 49.08                &46.58&18.36&19.04 & 35.43                 \\ \cline{2-18}
                           &                              &                              & Static                 & 48.38                & 49.96                & 42.89                & 45.88                & \cellcolor[HTML]{C0C0C0}41.74 & \cellcolor[HTML]{C0C0C0}49.88 & 49.96                         & 49.95                         & 49.96                & 45.03                &47.05&45.81&49.96 & 47.36                 \\
                           & \multirow{-2}{*}{$\times$} & \multirow{-2}{*}{\checkmark}   & $\Delta \Delta^{2}$    & 47.83                & 40.07                & 49.57                & 49.94                & \cellcolor[HTML]{C0C0C0}30.61  & \cellcolor[HTML]{C0C0C0}31.82 & 27.10                         & 28.22                         & 42.81                & 48.44               &46.44&31.22&27.66  & 39.64                 \\ \cline{2-18}
                           &                              &                              & Static                 & 49.99                & 49.92                & 40.11                & 49.99                & 45.22                         & 49.96                         & \cellcolor[HTML]{C0C0C0}49.99 & \cellcolor[HTML]{C0C0C0}47.36 & 50.00                & 50.00                &48.34&47.59&48.68 & 48.25        \\
\multirow{-6}{*}{CQCC}     & \multirow{-2}{*}{\checkmark}   & \multirow{-2}{*}{$\times$} & $\Delta \Delta^{2}$    & 44.44                & 46.12                & 42.62                & 49.47                & 9.87                        & 13.80                         & \cellcolor[HTML]{C0C0C0}21.89 & \cellcolor[HTML]{C0C0C0}23.71 & 49.95                & 49.96                &47.09&11.84&22.80 & 35.18                 \\ \hline
\end{tabular}
\vspace{-0.4cm}
\end{table*}
\vspace{-13pt}
\section{Results and Analysis}
\vspace{-10pt}
\subsection{BTAS 2016}
\vspace{-5pt}
We first conduct an experiment on BTAS 2016 replay spoofing development dataset to investigate the effects of different training data. The aim is to learn the system's ability to detect spoofed signals generated by various spoofing algorithms that are not incorporated in the training phase. Overall performance evaluation  results on eight replay attacks (R1-R8), obtained using conventional MFCC and proposed CQCC feature based countermeasures are reported in Table \ref{dev}. Due to attack dependency, the performance degrades drastically when one attack type is excluded from training the models and when the system is confronted with the similar type of attack in the system assessment process. We also observe that including direct replay in training helps for both SS and VC but not vice versa. This can be justified by the fact that SS and VC attacks are different whereas replay attacks have high similarity with genuine speech in terms of frequency components and formant trajectories \cite{wu2015spoofing}. Consequently, the replay speech characteristics of natural signal cannot be captured properly when they are eliminated from training the models. Interestingly, the static spectral features lead to promising recognition accuracy as opposed to  their dynamic counterparts. This is in contrast to previous studies \cite{sahidullah2015comparison, todisconew}.

We perform further experiments for only static features on the evaluation dataset. The overall and individual results are reported in Table \ref{eval}. CQCC feature yields superior result in all training conditions. This probably can be explained by the fact that CQCC feature provides higher resolution in lower and higher frequency regions that reflects better human perception system. Thus, they contribute better ability to capture replay characteristics while the models are trained by entire or a specific type of attacks. Furthermore, the pattern in EER values represents similar nature when features from unknown spoofing classes appear in the evaluation phase. It is worthwhile to mention that overall performance is compromised throughout all generalization systems for such unknown attacks (R9-R10). Comparing CQCC feature with MFCC feature, the CQCC feature outperforms other systems reported in \cite{korshunovoverview} with an average EER of 0.76 \%.

\vspace{-12pt}
\subsection{ASVspoof 2015}
\vspace{-5pt}
The results of generalized systems on ASVspoof 2015 synthetic spoofing database are reported in Table \ref{asvspoof_eval}. We train the countermeasure with both and either of the SS and VC attacks. In this study, we find that the dynamic coefficients provide superior performance in detecting synthetic spoofed signals. The results show a large amount of deterioration in performance when a particular attack is not considered in training. Although both MFCC and CQCC features give poor performance, CQCC feature leads to better performance across all cases of generalization scenarios. It is also interesting that for a particular case of generalization (where SS type attack is only used for training), static features give higher recognition accuracy than dynamic features for S1 and S10 attacks. We also notice that best performance for a specific attack is obtained if data from the specific attack type is used in training.
\vspace{-12pt}
\subsection{Cross-corpora Evaluation}
\vspace{-5pt}
The goal of this study has been to check cross-corpora vulnerability in a similar attack dependency framework where ASVspoof 2015 synthetic data is used to model the countermeasure and system performance is evaluated on BTAS 2016 replay evaluation dataset. The cross-corpora evaluations are shown in Table \ref{crosscorpora_eval}. The overall performance is poor as SS and VC data of BTAS 2016 test set consist of replayed version of SS and VC attacks as oppose to the ASVspoof database.  An interesting observation is that although replay attacks show poor recognition accuracy, dynamic features convey more distinct information in case of SS and VC attacks. It seems reasonable given that SS and VC based spoofed data are better modeled through dynamic characteristics. This, in turn, enhances the recognition accuracy while detecting replay version of VC and SS spoofed samples. Conventional MFCC feature proves to be more efficient in cross-corpora evaluation, but the performance of unknown attacks is poor for both features.
\vspace{-10pt}
\section{Conclusions}
\vspace{-5pt}
This work presents first analysis of spoofing countermeasures for attack dependency and generalization. A detailed study on BTAS 2016 with extensive experiments reveals that direct replay data have better generalization capability than SS and VC based replayed data. Results on ASVspoof 2015 demonstrates that VC spoofed data in training can better represent the attack space. The cross-corpora evaluation performance is very poor due to lack of suitable data in training. Our study on both the databases also indicates that both static and dynamic parts of spectral features are useful for detecting spoofing attacks in generalized sense.


\vspace{-0.6em}
\scriptsize
\section{Acknowledgment}
\vspace{-0.6em}
This work is partially supported by Indian Space Research Organization (ISRO), Government of India. The paper reflects some results from the OCTAVE Project (\#647850), funded by the Research European Agency (REA) of the European Commission, in its framework programme Horizon 2020. The views expressed in this paper are those of the authors and do not engage any official position on the European Commission.

 \newpage
 \clearpage

\ninept
\bibliographystyle{IEEEtran}

\bibliography{latexbib}

\end{document}